\pdfoutput=1
\documentclass[11pt]{article}
\usepackage[margin=1.15in]{geometry}
\usepackage{amsmath,amssymb,amsthm}
\usepackage{graphicx}
\usepackage{booktabs}
\usepackage{nicefrac}
\usepackage[colorlinks=true,linkcolor=black,citecolor=black,urlcolor=blue]{hyperref}
\usepackage{microtype}

\theoremstyle{plain}
\newtheorem{proposition}{Proposition}
\newtheorem{corollary}[proposition]{Corollary}
\theoremstyle{remark}
\newtheorem{remark}[proposition]{Remark}
\newtheorem*{assumption*}{Assumption (Common Cohort)}
\newtheorem*{assumptionb*}{Assumption (Net Restraint)}
\newtheorem*{assumptionc*}{Assumption (Interior Grading)}

\newcommand{\gs}{g^{*}}
\newcommand{\E}{\mathbb{E}}
\newcommand{\Var}{\mathrm{Var}}

\title{\bf Degrading the Grade: Grading Variance as Evidence of Shirking}
\author{Thayer Morrill\thanks{Department of Economics, North Carolina State University.  Email: \texttt{tsmorril@ncsu.edu}.  The author has no conflicts of interest to declare.  The author used Claude (Anthropic) to edit and proofread the text, to check the proofs and algebra, and to check the references, and takes full responsibility for the content of the paper.}}
\date{November 2026}

\begin{document}
\maketitle
\vspace{-2.5em}

\begin{abstract}
\noindent Variation in grading standards across instructors who teach the same students is evidence of a moral hazard problem, whether or not grades are inflated.  The optimal standard is the same for every section of a course, yet grades vary widely across sections.  We present a principal-agent model in which instructors who know the optimal standard still depart from it, leniently or harshly, depending on their incentives.  Penalizing deviations from the department average leaves the average unchanged but compresses grading variance, which raises student welfare.  Disclosing the same information to students instead increases grading variance.

\end{abstract}

\section{Introduction}

In October 2025, Harvard's Office of Undergraduate Education reported that 60\% of the grades it awards are now A's, compared to 24\% in 2005 (Claybaugh, 2025).\footnote{The Harvard Crimson reported that the median grade point average
of a graduating class had reached $3.83$ (Church and Srivastava, 2025).}  The report described the current practices as ``failing to perform the key functions of grading,'' and the Harvard Crimson described it as failing (Church and Srivastava, 2025).  The reaction, at Harvard and
elsewhere, has been to focus on the rise in the grade \textsl{average}.  Harvard's faculty later voted to cap the number of A's in a course (Lee, 2026), while others have studied constraining the grade point average of every course (Ahn, Arcidiacono, Hopson, and Thomas, 2024).

We argue that university administration should focus on grading \textsl{variance}.  It is not obvious that a rising average grade is a
distortion at all. Just as consumers become comfortable with new prices, employers become comfortable with new norms.  What degrades grades as a signal is when the same grade means different things.  Further, there is no reason the same grade distribution needs to be optimal in different places and at different times.  A grading standard has competing effects on student effort. If an A is too easy to obtain, then it does not require work.  If an A is too hard to obtain, then the effort is too costly for the reward (Costrell, 1994; Dubey and Geanakoplos, 2010).  The optimal policy balances how students respond to encouragement versus
penalty, and there is no reason to expect the same answer at Harvard as at North
Carolina State University (NC State), nor the same answer at Harvard in 2025 as in 2005. Whether the
optimal distribution has drifted upward over two decades is an empirical
question and is perhaps unknowable.

There is, however, one comparison in which the optimal distribution cannot
plausibly differ.  Two sections of the same course, taught in the same term, drawing from
the same cohort, should be graded to the same standard. At NC State, in the same class (Principles of Microeconomics) in the same semester (spring 2025), one instructor awarded 17\% A's while another awarded A's to more than 80\% of students. This variance is remarkable.  What does an A in Principles of Microeconomics from NC State signal?  It depends on the section!

We argue that an institution should be concerned about the variance whether or not it is concerned about the mean.  To explain where the variance comes from, we present a simple principal--agent
model. A substantial literature studies what the optimal grading policy is; that
is not our question, and we take the optimal policy as exogenous. Our question is
why an instructor who \emph{knows} the optimal policy would nonetheless depart
from it. The tension in our model is natural.  Accurately grading an exam is costly.  It is less costly to grade questions that everyone gets right or wrong.  Judgment calls are what make grading difficult, and for simplicity, we assume that the most difficult grading occurs at the optimal distribution.  Therefore, an instructor can reduce her effort by either grading generously or strictly.

Lenient grading has tradeoffs.  We assume that a lenient grader receives better evaluations and attracts more students.  A harsh standard results in the reverse.  Instructors differ in how they evaluate these tradeoffs.  To some instructors, bad evaluations are costly, and to others, they are not.  Similarly, an instructor may view larger enrollment as a burden or as a benefit.  The result is that a department of instructors who all know the optimal standard and teach the same students will nonetheless vary in the grade distributions they assign.  Therefore, in our model, the department mean can look fine even when there is a moral hazard problem; it is the variance in grading standards that reveals it.  

In our model, the department mean corresponds to the average of the instructors' biases relative to their restraint.  In particular, for a symmetric distribution of biases that is independent of restraint, the department average will be optimal; however, almost no individual instructor will employ the optimal distribution.  We assume that student welfare is maximized at the optimal grade distribution.  The variation in grading policies results in a loss of student welfare.  However, a principal observing only the departmental average would not realize there is a problem.     Note that the loss in student welfare is only part of the social cost of grading variance, as it also degrades the quality of a signal that a grade provides.

Finally, we consider a natural policy intervention.  We assume that the principal does not know the optimal grade distribution.  Nonetheless, she observes the departmental average.  We consider the principal imposing a cost on the instructor that is an increasing function of the distance between the instructor's average and the departmental average.  We show that this policy does not change the department average, but it does compress the dispersion of grading policies.  Therefore, although it has no effect on grade inflation, it results in an unambiguous improvement in student welfare due to reduced grading variance.  A natural worry is that such a rule invites the instructors to coordinate, for example on giving every student an A.  We show that this cannot happen as long as small deviations from the average are not penalized harshly, and that it is each instructor's own restraint, not the existence of an optimal standard, that rules it out.

\paragraph{Related literature.} The paper sits at the intersection of a number of literatures.  First is the literature on optimal grading policy.   Theoretical papers have considered the optimal grading policy for eliciting effort (Costrell, 1994; Dubey and Geanakoplos, 2010) or as a signaling device (Boleslavsky and Cotton, 2015; Chan, Li, and Suen, 2007; Popov and Bernhardt, 2013). We take the optimal standard as
exogenous and ask about dispersion around it.  Ostrovsky and Schwarz (2010) note that inconsistent standards across instructors add noise to transcripts and that a school may tolerate this noise because it helps its students' job placement.  Our welfare criterion is student welfare rather than placement, and in our model, the dispersion is chosen by the instructors, not by the school.

Our paper is most directly related to the literature on instructors' grading incentives.  Economists have long modeled grading as a choice instructors make in their own interest.  In Dickson (1984), lenient grading raises enrollment, which brings job security but also more work.  Love and Kotchen (2010) model a professor who trades better evaluations against student effort and research time, Franz (2010) has professors inflate grades to avoid being pestered by students, and Correa (2001) shows that competition for students drives standards down.  Achen and Courant (2009) argue that departments use grades to manage their enrollments.  What is new in our model is that instructors err in both directions around a known optimum, and that the effort saving can push an instructor either way.  Bonesr{\o}nning (1999) documents large differences across Norwegian schools in the grades teachers give for the same external exam score, and relates them to teacher characteristics.  In his model the principal imposes a cost on teachers who depart from her preferred standard; our chair needs no target and penalizes departures from the realized average.

The closest paper to ours is Gans and Kominers (2026). They are concerned with what inferences can be made about a student from a transcript.  Because courses differ in both the difficulty and composition of students, a grade is not comparable
across courses.  They show how a full transcript can recover comparable student signals.  Like our paper, they consider instructors' incentives.  In fact, our instructor's payoff function shares three of its four components with theirs: a
private return to leniency, an enrollment motive, and a quadratic cost of
departing from a norm.  Their parameters are common across instructors, so their equilibrium is symmetric, and dispersion is zero by construction.  They show that with a better transcript, an instructor can no longer attract students by being a more lenient grader, thus eliminating what they call the competitive enrollment component of grade inflation.  Our instructor is a heterogeneous version of theirs, with a linear rather than strategic enrollment motive and an added cost of grading effort.  Our objectives are different.  Their objective is to ``deflate'' grade inflation, while ours is to highlight the cost introduced by grading variance.

Ahn, Arcidiacono, Hopson, and Thomas (2024) estimate an equilibrium model in which grading policies respond to course demand.  Their professors also differ in how they grade, but because each has her own ideal average grade, not because they depart from a common optimum that they all know.  Their central counterfactual is a policy requiring that every course have the same average grade, a B.  Our intervention is different from theirs in two key ways.  Their intervention is a constraint while ours is a penalty.  Further, theirs fixes the average to a chosen target, while ours requires no target.  Ours penalizes deviations from the realized average, which is why it does not change the mean.  Gregor (2021) studies grading competition when students shop across electives, and the school caps the mean score.

Our proposed intervention --- penalizing each
agent by her distance from the average of the others --- has long been considered in the IO literature on relative performance evaluation.  Formally, it is yardstick regulation (Shleifer, 1985).  Relative schemes such as tournaments (Lazear and Rosen,
1981) have been used in the classical relative-performance literature pioneered by Holmstr\"om (1982) and Green and Stokey (1983).  In the classical relative-performance literature, the objective is to filter out noise.  Our point is that compression can be beneficial regardless of whether the average is at the optimum.\footnote{There is also a long literature in which agents are punished for deviating from a group average resulting in bunching.   Bernheim (1994) obtains conformity
around an endogenous norm from a status motive.  Kandel and Lazear (1992) and
Fischer and Huddart (2008) build peer pressure and endogenous norms into
organizational contracting.  In Fischer and Huddart (2008) and Glaeser and Scheinkman (2003), the weight placed on
conforming is taken away from the agent's own objective, and this is what moves
the population mean.} 


Finally, the paper is related to work on agency problems on the evaluator's
side of subjective performance evaluation (Prendergast and Topel, 1996; MacLeod,
2003).  In our context, an instructor is an evaluator, and grade dispersion is the result of moral
hazard.

\paragraph{Outline.} Section 2 sets up the environment, Section 3 states the
student benchmark and fixes the welfare criterion, and Section 4 presents the
instructor's problem. Section 5 solves it: an instructor's departure from the
optimum is her bias divided by her net restraint, and a department's grading
distribution is therefore the distribution of its incentive problem. It ends with
the welfare decomposition into grade inflation and grading variance. Section 6 models the
chair's intervention as a game and establishes its three properties --- uniqueness,
compression, and neutrality with respect to the section average --- and then shows
that the same information disclosed to
students instead has the opposite effect. Section 7 collects testable
implications, Section 8 discusses what is and is not doing the work, and Section 9 concludes. The
appendix relaxes the assumptions of the main text.

\section{Setup}

A department consists of a unit mass of instructors indexed by $i$. Each teaches
one section of a common course.  We assume that enrollment may respond to grading (Section 4), but the composition of a section does not.

Instructor $i$ chooses a \emph{grade distribution}, summarized by a single-dimensional variable
$g_i\in[0,1]$ (for example, the share of students who receive an A).  A higher $g$ means a more
lenient grading standard. Nothing in the argument depends on the particular summary chosen, but we use a one-dimensional variable for clarity and simplicity.

\begin{assumption*}
All instructors draw from the same cohort, so the optimal standard $\gs$
is common across $i$.
\end{assumption*}

\noindent Our assumption is deliberately restrictive.  We wish to explain where grading variance might arise, even when it is clear there should be none.  We are not trying to explain why a principles course has a different distribution from a senior elective; this seems natural.  Similarly, we are not trying to explain why courses across different majors might have different distributions; this also seems natural.  Rather, we are trying to explain why two different sections of the exact same principles course, in the same semester, might have different distributions:  this seems unnatural.\footnote{Such differences are well documented.  Bonesr{\o}nning (1999) finds large differences across Norwegian schools in the grades teachers give for the same external exam score, and McKenzie and Allensworth (2025) find that differences across teachers in how they grade are stable from year to year.  NC State publishes the grade distribution of every section of every course, by instructor, through its Gradient tool.  Figure~\ref{fig:ncstate} shows Principles of Microeconomics, and we have found large differences across sections taught in the same term in every large multi-section course we have examined.}

\section{Welfare}

Our principal (such as a department chair or a dean) has the objective of maximizing \emph{student} welfare.  Specifically, the principal considers neither the benefit to the instructor from shirking nor the cost to the instructor if the principal imposes a punishment.  We do not explicitly model student welfare from a grade distribution.  We assume that there is an optimal grading policy $\gs\in(0,1)$\footnote{The evidence is consistent with an interior optimum.  Lenient grading lowers effort: holding the course and the instructor fixed, Babcock (2010) finds that study time is about half as large when the class expects an A as when it expects a C.  Students of teachers with higher grading standards do better in later math courses (Gershenson, Holt, and Tyner, 2024), and students taught by stricter graders earn more later (Denning, Nesbit, Pope, and Warnick, 2026).  Harsher is not always better, however.  Holding a teacher's effectiveness fixed, McKenzie and Allensworth (2025) find that more lenient grading slightly improves students' later outcomes.} and that student welfare from grading policy $g$ is: 

\begin{equation}\label{eq:W}
W(g) \;=\; -\tfrac{\kappa}{2}\,(g-\gs)^{2}, \qquad \kappa>0 .
\end{equation}

We assume this simple form for clarity.  The welfare decomposition in Section~\ref{sec:welfare} uses the quadratic form, but our welfare result for the intervention holds for any concave $W$ (Proposition~\ref{prop:ssd}).  We take $\gs$ as a primitive; its existence and location are the subject of the optimal-grading literature (Costrell, 1994; Dubey and Geanakoplos, 2010).  Our point is that even if $\gs$ is known to everyone, it will not be chosen.

As we consider an abstract grading policy, we assume for simplicity that the optimal grading distribution is at the midpoint, which is where grading is most costly (Section~4).  We therefore set
\[
\gs \;=\; \tfrac{1}{2}
\]
throughout.\footnote{With a general $\gs$ every deviation below acquires one
additional additive term, $-\theta_i(1-2\gs)/\lambda_i$, which tilts all
instructors toward whichever boundary is nearer. This shifts the department average away from $\gs$ even when bias is symmetric, and it means the cost of grading can affect the direction of an instructor's deviation, not only its size.  It does not change our results on dispersion, the neutrality of the intervention, or its effect on student welfare, which depend only on deviations from the department average.  We suppress it for readability.}

\section{The instructor's problem}

Instructor $i$'s payoff has four parts.

\paragraph{Conscientiousness.} The instructor benefits from doing her job well.  She puts weight $\mu_i>0$ on doing the
job properly, contributing $-\tfrac{\mu_i}{2}(g-\gs)^{2}$. We interpret $\mu_i$ as
conscientiousness, professional norm, or whatever institutional pressure
already exists. 

\paragraph{Teaching evaluations.} We assume that more generous graders receive better teaching evaluations.  Specifically,  evaluations are
increasing in $g$, and instructor $i$ places weight $\alpha_i\ge 0$ on them,
contributing $\alpha_i g$. The weight is heterogeneous because teaching evaluations have a different impact on different instructors.  Evaluations matter more to an instructor going up for tenure than they do to a senior colleague with a strong research record.  The assumption that leniency is rewarded is well supported.  Teachers who give higher grades receive better evaluations, while more effective teachers receive worse ones (Braga, Paccagnella, and Pellizzari, 2014), and evaluations are about three times as strongly correlated with an instructor's leniency as with her value added (Warnick, Light, and Yim, 2026).  Carrell and West (2010) find that evaluations reward teaching that raises current performance at the expense of later learning.

\paragraph{Enrollment.}  The instructor's grading policy affects her enrollment.  A lenient grading policy attracts students: enrollment is
$n(g)=n_0+\delta g$ with $\delta>0$, where $\delta$ measures how responsive
students are to grading practices. Instructor $i$'s marginal valuation of a
student is $-\gamma_i$, contributing $-\gamma_i \delta g$. We allow
$\gamma_i \gtrless 0$: some instructors prefer a small class as it takes less time to grade while others may need enrollment to keep an elective alive.  Students do choose courses partly by how they are graded (Sabot and Wakeman-Linn, 1991).  Ahn, Arcidiacono, Hopson, and Thomas (2024) estimate a model with the same features using data from the University of Kentucky: students' course choices respond to grading, and each professor has her own ideal enrollment, so that professors raise grades to attract students when demand is low and lower grades to deter students when demand is high.

\paragraph{Grading effort.} Grading is costly but not uniformly so.  It is easier to grade a question that everyone gets correct, or that everyone gets wrong.  Therefore, we assume the cost of grading is proportional to how evenly
the class splits,
\begin{equation}
c_i(g) \;=\; \theta_i\, g(1-g), \qquad \theta_i\ge 0,
\end{equation}
which vanishes at both corners and peaks at $g=\nicefrac{1}{2}$. Note that $g(1-g)$ is exactly the
variance of whether a student in her class receives an A.

More generally, if students' performance has density $f$ and the instructor sets a
cutoff $\tau$ with $g=1-F(\tau)$, the marginal-judgment cost is proportional to
$f(\tau)$.  For a single-peaked density, this cost is largest in the middle and small at the extremes.  Our qualitative results extend to $c_i(g)=\theta_i\psi(g)$ with $\psi$ strictly concave, symmetric about $\nicefrac{1}{2}$, and $\psi(0)=\psi(1)=0$, but at the cost of the closed forms.  We assume the simple, quadratic form for clarity.

\paragraph{The instructor's objective.} Collecting terms, we define the \emph{instructor bias} as
\begin{equation}
b_i \;\equiv\; \alpha_i - \delta\gamma_i. 
\end{equation}

The private trade-offs of grading leniency are better evaluations and a larger (or smaller) class.  The instructor $i$ solves

\begin{equation}\label{eq:obj}
\max_{g\in[0,1]}\;\;
U_i(g) \;=\; -\frac{\mu_i}{2}(g-\gs)^{2} \;+\; b_i\,g \;-\; \theta_i\, g(1-g).
\end{equation}

We have deliberately made the trade-offs of leniency --- better evaluations and a larger (or smaller) class ---
linear.  As a result, they collapse into the single sufficient statistic $b_i$.
All instructors care about evaluations and class size, but what matters for behavior is which of these two dominates.

\paragraph{Net restraint.} Two of the three terms in \eqref{eq:obj} are
quadratic. Conscientiousness pulls toward $\gs$ with curvature $\mu_i$; the effort
saving pushes away from the middle with curvature $2\theta_i$. Only their
difference matters, and we call it the instructor's \emph{net restraint}:
\begin{equation}\label{eq:lambda}
\lambda_i \;\equiv\; \mu_i - 2\theta_i .
\end{equation}

This is the only combination of $\mu_i$ and $\theta_i$ that appears anywhere
below (given $\gs=\nicefrac{1}{2}$).  To ensure that each instructor has a unique preferred standard, we assume 

\begin{assumptionb*}
$\lambda_i\ge\underline{\lambda}>0$ for every instructor.
\end{assumptionb*}

\noindent In words, each instructor cares more about doing the job right (or her professional reputation) than what careful grading costs her.  The assumption makes $U_i$ strictly concave, and the uniform bound keeps $\E[1/\lambda_i]$, and hence the average grading standard, finite.  Table~\ref{tab:notation} summarizes the notation.

\begin{table}[t]
\centering
\small
\begin{tabular}{@{}lll@{}}
\toprule
Parameter & Meaning & Role \\
\midrule
$\gs$ & optimal standard & common to all instructors \\
$\mu_i$ & weight on doing the job properly & damps the distortion \\
$\alpha_i$ & sensitivity to teaching evaluations & pulls toward leniency \\
$\gamma_i$ & marginal cost of an extra student & pulls toward harshness if $>0$ \\
$\delta$ & responsiveness of enrollment to leniency & scales the enrollment channel \\
$\theta_i$ & cost of grading effort & amplifies the distortion (given $\gs=\nicefrac12$) \\
$b_i=\alpha_i-\delta\gamma_i$ & instructor bias & \emph{signs} the distortion \\
$\lambda_i=\mu_i-2\theta_i$ & net restraint & \emph{scales} the distortion \\
\bottomrule
\end{tabular}
\caption{Notation.}
\label{tab:notation}
\end{table}

\section{Equilibrium grading standards}

\subsection{One instructor}

Differentiate \eqref{eq:obj} and collect terms.  We are left with a simple first-order condition:

\begin{equation}\label{eq:foc}
U_i'(g) \;=\; \underbrace{-\lambda_i\left(g-\tfrac{1}{2}\right)}_{\text{marginal cost of deviating}}
\;+\; \underbrace{b_i}_{\text{private pull}},
\qquad
U_i''(g) = -\lambda_i < 0 .
\end{equation}

The marginal cost of moving away from the optimal grading policy rises by $\lambda_i$ per unit of distance.  The marginal benefit is $b_i$, her private pull.

In large classes, we do not expect any instructor to give 100\% or 0\% A's.  Interior solutions follow from a simple assumption.

\begin{assumptionc*}
$|b_i| < \lambda_i/2$ for every instructor.
\end{assumptionc*}

\noindent In words, an instructor's net restraint is more than twice her private pull.  In the appendix, we drop this assumption: the balance result below survives, but the observed variance then understates the dispersion of incentives.

\begin{proposition}[Grading standards]\label{prop:one}
Instructor $i$ chooses
\begin{equation}\label{eq:dev}
g_i \;=\; \tfrac{1}{2} \;+\; \frac{b_i}{\lambda_i}.
\end{equation}
\end{proposition}

\begin{proof}
$U_i$ is strictly concave by Assumption (Net Restraint), so the first-order
condition is necessary and sufficient; setting \eqref{eq:foc} to zero gives
\eqref{eq:dev}, and Assumption (Interior Grading) places it inside $[0,1]$.
\end{proof}

\noindent \textbf{Deviation equals bias divided by restraint.}  This is the basic point of our model.  Even in a setting where (1) all instructors teach the same students, (2) there is an optimal grade distribution, and (3) this optimal distribution is known to everyone, there will be variance in grading policies.  The variance is a consequence of misaligned incentives.  

We highlight the key takeaways:
\begin{itemize}
	\item An instructor is lenient or harsh, depending on the sign of $b$.  An instructor who prioritizes evaluations, or who wants a larger class, will be lenient, and an instructor who dislikes large classes more than she values evaluations will be harsh.
	\item As $\lambda_i=\mu_i-2\theta_i$, the size of the deviation, $|g_i-\nicefrac{1}{2}|$, is decreasing in her professionalism, $\mu_i$, and increasing in $\theta_i$.  To emphasize, given $\gs=\nicefrac{1}{2}$, the cost of grading does not determine whether an instructor is harsh or lenient, but rather, this amplifies the size of her deviation.
\end{itemize}

\begin{figure}[htbp]
\centering
\includegraphics[width=\textwidth]{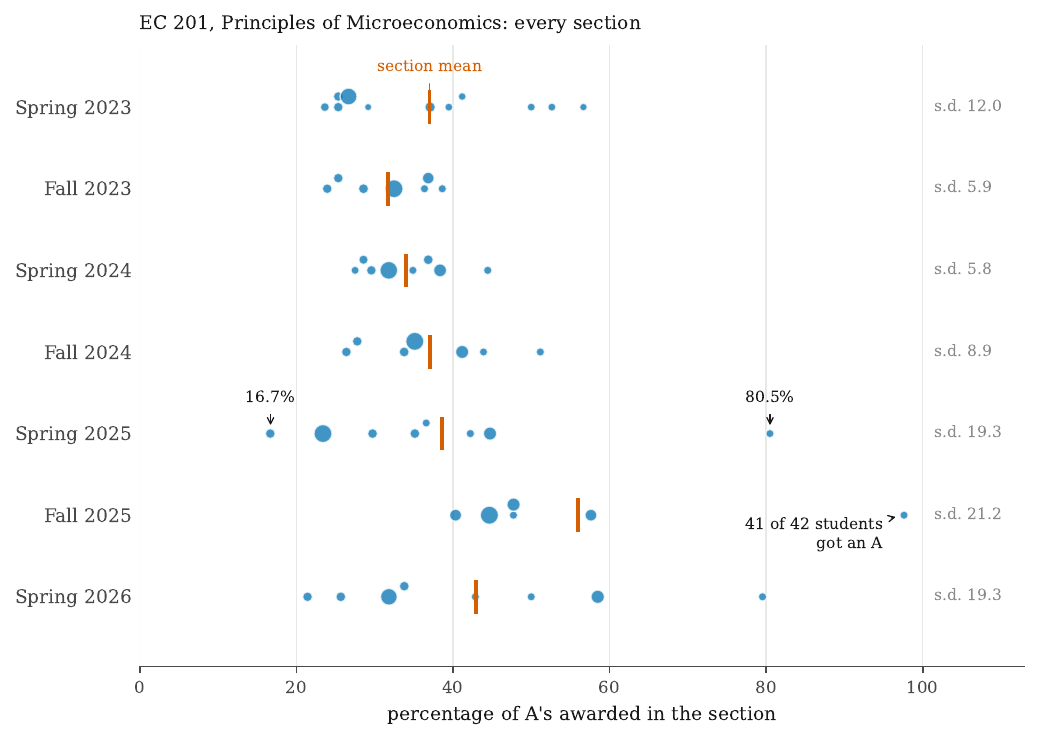}
\caption{Share of A's awarded, by section, in EC 201 (Principles of
Microeconomics) at NC State, every term from spring 2023
through spring 2026. Each circle is one section; circle area is proportional to the
number of letter grades awarded, and the denominator is A--F grades excluding
withdrawals. Sections with fewer than ten letter grades are omitted. Dispersion
across sections of the same course in the same term rises sharply in 2025: in
spring 2025 one section awarded A's to $16.7$ percent of its students while
another awarded them to $80.5$ percent, and in fall 2025 one section awarded an A
to $41$ of its $42$ students.  The dispersion persists in spring 2026, when sections ranged from $21.4$ to $79.5$ percent A's. Term-level standard deviations are reported at the
right; note that the number of sections varies across terms, from eleven in spring
2023 to six in fall 2025. Source: NC State Gradient.}
\label{fig:ncstate}
\end{figure}

Figure~\ref{fig:ncstate} gives the percentage of A's given in different sections
of Principles of Microeconomics at NC State in every term from spring 2023 through spring 2026. To emphasize, these are different sections of the exact same course
offered in the same term.

Our model says that an instructor whose cost of grading is large relative to her professional concern will give nearly all A's or essentially none, and that which she gives is determined by her bias.  An instructor who is sensitive to teaching evaluations (such as a graduate student) will give almost all A's, while an instructor who is not and who would rather teach a smaller class (such as a senior colleague near retirement) will give almost none.

\begin{figure}[htbp]
\centering
\includegraphics[width=\textwidth]{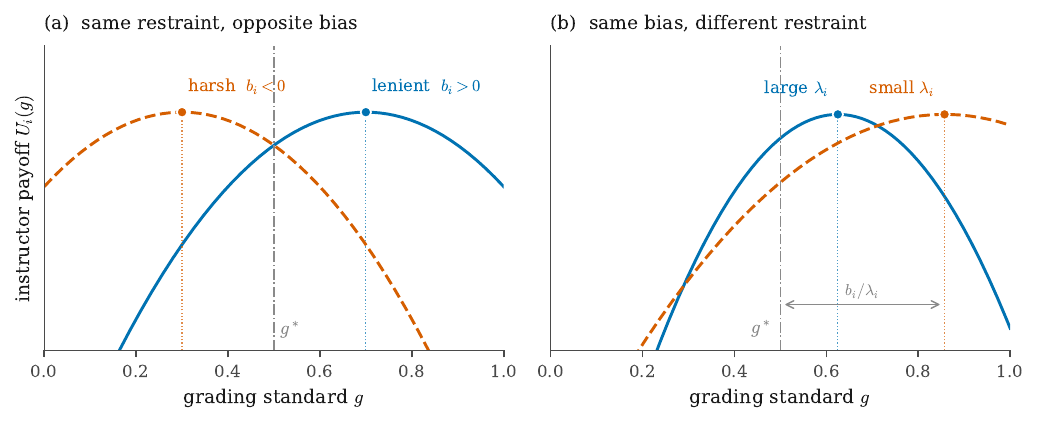}
\caption{Every instructor's objective is a concave function of the grading
standard, peaked at $\nicefrac12 + b_i/\lambda_i$. Panel (a): two instructors with
the same restraint and opposite biases, both knowing $\gs$ and facing the same
cohort, choose standards on opposite sides of it. Panel (b): two instructors with
the same bias but different restraint depart from the optimum by different
amounts.}
\end{figure}

\subsection{The department average}

As we know the distribution each instructor will choose from Proposition~\ref{prop:one}, it is straightforward to calculate what the average grade policy will be for a department.   Let $F$ denote the distribution of $\nicefrac12 + b_i/\lambda_i$ across the department.  Then the department's grade policy is distributed according to $F$.

\begin{proposition}[The department's grading distribution]\label{prop:dept}
The department's average grading standard is the
average of $F$, and the variance of grading standards across sections is the variance
of $F$.
\end{proposition}

\begin{proof}
This is immediate from Proposition~\ref{prop:one}.  As each instructor chooses her preferred standard (since each instructor's preferred standard is in the interior), the department's average is the expected value of $F$.  Similarly, the distribution of grading standards follows the distribution of $F$.
\end{proof}

\noindent The key point is that the distribution of grading policies in a department coincides with the distribution of the incentive problem.  In other words, by observing the distribution of grading policies for a department, we observe the distribution of the incentive problem: each instructor's bias relative to her restraint. 

In this model, there is no \textsl{a priori} expectation of grade inflation.  An immediate consequence of Proposition~\ref{prop:dept} is that if the distribution of $b_i/\lambda_i$ is symmetric about zero (for example, if bias is symmetric and independent of restraint), the department average will be the optimal grading policy.  However, if instructors are incentivized to have high evaluations, then this bias will induce grade inflation.

\begin{corollary}\label{prop:balanced}
Suppose the distribution of $b_i/\lambda_i$ is symmetric about zero, as it is when bias is symmetric about zero and independent of restraint.  Then the department's average grading standard is
exactly $\gs$.
\end{corollary}

\begin{proof}
The deviations $b_i/\lambda_i$ are symmetric about zero, so they
average to zero, and by Proposition~\ref{prop:dept} the average standard is $\gs$.
\end{proof}

\noindent Our interpretation of Corollary \ref{prop:balanced} is that university policy should be focused at least as much on grading variance as grade inflation.  In our model, when bias relative to restraint is symmetric, the department has the optimal average grading policy with almost no individual instructor grading optimally.  Such a department has a welfare loss, yet we cannot discern this from only looking at the average.

\subsection{Welfare loss}\label{sec:welfare}

Here we summarize the welfare loss in our principal-agent problem.  We show that the average grading policy can conceal substantial welfare losses for students due to variation in grading policies.  Let the parameters $(b_i,\theta_i,\mu_i)$ be jointly distributed across the department.

We emphasize, as in Section~3, that our measure of welfare is student welfare.  Specifically, we do not include the ``benefit'' the instructor receives from shirking.  We denote student welfare by $W$ and write the
student loss as $\mathcal{L}=-\E[W(g_i)]=\tfrac{\kappa}{2}\,\E[(g_i-\gs)^2]$.  This splits into two pieces.

\begin{proposition}[Bias--variance decomposition of grading distortion]\label{prop:four}
\begin{equation}\label{eq:bv}
\mathcal{L} \;=\; \frac{\kappa}{2}\Big[\underbrace{\big(\E[g_i]-\gs\big)^{2}}_{\text{grade inflation}}
\;+\; \underbrace{\Var(g_i)}_{\text{grade dispersion}}\Big].
\end{equation}
\end{proposition}

\begin{proof}
Add and subtract $\E[g_i]$ inside the square and expand; the cross term vanishes.
\end{proof}

\begin{corollary}[When the whole loss is variance]\label{prop:allvariance}
Suppose restraint is common across instructors, $\lambda_i\equiv\lambda$, and bias
is symmetric about zero. Then the department exhibits no grade inflation, and the
entire loss is grading variance:
\[
\mathcal{L} \;=\; \frac{\kappa}{2}\,\Var(g_i) \;=\; \frac{\kappa\,\sigma_b^{2}}{2\lambda^{2}},
\qquad \sigma_b^{2}=\Var(b_i).
\]
\end{corollary}

\begin{proof}
By Corollary~\ref{prop:balanced} symmetric bias gives $\E[g_i]=\gs$, so the first
term in \eqref{eq:bv} is zero. By Proposition~\ref{prop:dept} the second term is
$\Var\!\left(b_i/\lambda\right)=\sigma_b^{2}/\lambda^{2}$.
\end{proof}

\noindent There has been a great deal of attention to grade inflation.  Equation \eqref{eq:bv} is consistent with this.  Grade inflation may indicate welfare loss.  However, grade inflation is not proof of welfare loss.  The optimal grade distribution need not be fixed across time.  For example, due to changes in parenting styles, a generation of students may be better incentivized by positive reinforcement than a previous generation was.  However, the point of this paper is that grading variance always induces welfare loss.

\section{Intervention}\label{sec:intervention}

There is a natural solution to this principal-agent problem.  Here, we interpret the principal as someone with authority over the instructor (such as a chair, the dean, or a promotion committee).  The principal does not know the optimal distribution, but she does observe each instructor's distribution and therefore the department's average distribution.  A natural intervention is to punish an agent for deviating from the departmental mean.  We show that this intervention unambiguously increases student welfare.  We stress the word \emph{student}.   We have not considered the faculty's benefit from shirking in our welfare considerations.  But the chair's instrument is a punishment, and it is fair to ask whether the improvement in student welfare is worth the cost imposed on the faculty.  We address this in Appendix~\ref{app:costly}, where we show that a credible penalty is nearly costless in equilibrium.

Such an intervention induces a game, as each instructor's payoff depends on the other instructors' choices.  To simplify the analysis, we assume a continuum of agents so that each agent has a negligible impact on the department average.  We show in the Appendix that we reach the same conclusions when there is a finite number of agents.

\subsection{Punishing deviation from the norm}

The principal observes the department's grade distributions, computes the norm
$\bar g$, and punishes each instructor according to how far she sits from it.
There is no reason for that punishment to be linear. There is a wide range of ``punishments'' available to a chair, ranging from informal pressure to terminating employment.  We therefore allow a general penalty and assume
only that it is increasing at an increasing rate.

\begin{description}
\item[Penalty.] $P:\mathbb{R}\to\mathbb{R}_{+}$ with $P(0)=0$, $P(-d)=P(d)$,
$P$ convex, strictly increasing in $|d|$, and continuously differentiable except possibly at $0$. We write
$P'$ for its derivative, which is odd, and $\rho$ for a scalar intensity when
$P$ is taken from a parametric family such as $P(d)=\tfrac{\rho}{2}d^{2}$.
\item[Payoffs.] Instructor $i$ chooses $g_i$ to maximize
\begin{equation}\label{eq:game-obj}
U_i(g_i;\bar g) \;=\; -\frac{\mu_i}{2}(g_i-\gs)^{2} + b_i g_i - \theta_i g_i(1-g_i)
\;-\; P\!\left(g_i - \bar g\right),
\end{equation}
where $\bar g = \E[g_j]$ is the realized department norm.
\item[Equilibrium.] A Nash equilibrium is a profile $\{g_i\}$ and a norm $\bar g$
such that each $g_i$ maximizes \eqref{eq:game-obj} given $\bar g$, and
$\bar g = \E[g_i]$.
\end{description}

\noindent For this section we take restraint to be common across instructors,
$\lambda_i\equiv\lambda$, so that instructors differ only in their bias. The
Appendix allows restraint to vary.

Write $\hat g_i = \gs + b_i/\lambda$ for instructor $i$'s preferred standard
from Proposition~\ref{prop:one} --- what she chooses when no one is watching ---
and $d_i \equiv g_i - \bar g$ for her deviation from the norm. Substituting
$g_i = \bar g + d_i$ into the first-order condition for \eqref{eq:game-obj} and
rearranging gives the equation that organizes everything below:
\begin{equation}\label{eq:foc-game}
\underbrace{\lambda\, d_i + P'(d_i)}_{\textstyle \equiv\;\Phi(d_i)}
\;=\;\lambda\left(\hat g_i - \bar g\right).
\end{equation}
The left side is what it costs instructor $i$ at the margin to stand $d_i$ away
from her colleagues; the right side is what she gains at the margin by moving
toward where she would privately like to be. She stops when they balance.

\subsection{Three results}

\noindent The penalty may have a kink at zero.  We write $\rho_0 \equiv \lim_{d\downarrow 0}P'(d)\ge 0$ for the marginal penalty on the first unit of deviation, so that $\rho_0=0$ exactly when $P$ is differentiable at zero.  Let $\hat g_{\min}$ and $\hat g_{\max}$ denote the lowest and highest preferred standards in the department (the bounds of the support of $F$).

\begin{proposition}[Equilibrium]\label{prop:game}
Suppose $\rho_0 < \tfrac{\lambda}{2}\left(\hat g_{\max}-\hat g_{\min}\right)$, which holds whenever $P$ is differentiable at zero and the instructors' preferred standards are not all the same.  Then the game has a unique equilibrium. In it, every instructor's deviation from the
norm is smaller than it would have been without the intervention: for each $i$,
$d_i$ has the same sign as $\hat g_i - \bar g$ and $|d_i| \le |\hat g_i - \bar g|$, strictly whenever $\hat g_i\neq\bar g$.
\end{proposition}

\begin{proof}
Instructor $i$'s objective is strictly concave in $d_i$, so she has a unique best response $d_i(\bar g)$.  If $\lambda|\hat g_i-\bar g|\le\rho_0$, it is $d_i=0$.  Otherwise it is the unique solution of \eqref{eq:foc-game}, which is nonzero, has the sign of $\hat g_i-\bar g$, and is strictly decreasing in $\bar g$ because $\Phi$ is strictly increasing away from zero.  Hence $\E[d_i(\bar g)]$ is continuous and weakly decreasing in $\bar g$, and it is strictly decreasing at any $\bar g$ for which a positive mass of instructors has $\lambda|\hat g_i-\bar g|>\rho_0$.  Under our condition, this is every $\bar g$: either $\hat g_{\max}-\bar g$ or $\bar g-\hat g_{\min}$ is at least $\tfrac12(\hat g_{\max}-\hat g_{\min})>\rho_0/\lambda$, and by the definition of the support a positive mass of instructors lies beyond it.  $\E[d_i(\bar g)]$ is positive at $\bar g=\hat g_{\min}$ and negative at $\bar g=\hat g_{\max}$, so it crosses zero exactly once, which pins down $\bar g$. For the second claim, $\lambda d_i = \lambda(\hat g_i - \bar g) - P'(d_i)$ whenever $d_i\neq0$, and $P'(d_i)$ has
the same sign as $d_i$.
\end{proof}

\noindent For the rest of this section, we assume the condition of Proposition~\ref{prop:game} holds.  We return to what happens when it fails in Section~\ref{sec:coord}.

\begin{proposition}[The department average does not move]\label{prop:neutral}
If $F$ is symmetric about its mean, the intervention leaves the department's
average grading standard exactly where it was.
\end{proposition}

\begin{proof}
Let $\bar g^{0}$ denote the average of $F$ and conjecture $\bar g = \bar g^{0}$.
Then $\hat g_i - \bar g$ is symmetric about zero. Since $P'$ is odd, $\Phi$ is
odd and so is $\Phi^{-1}$; an odd function of a variable symmetric about zero is
symmetric about zero, so the $d_i$ are symmetric about zero and average to zero.
The conjecture therefore satisfies the equilibrium condition, and by
Proposition~\ref{prop:game} it is the only one.
\end{proof}

\noindent Symmetry is needed only because the penalty may be non-quadratic. For
the quadratic penalty $P(d)=\tfrac{\rho}{2}d^{2}$ the result holds for every $F$:
the first-order condition is then linear, giving
\begin{equation}\label{eq:weighted}
g_i \;=\; \frac{\lambda}{\lambda+\rho}\,\hat g_i \;+\; \frac{\rho}{\lambda+\rho}\,\bar g ,
\end{equation}
so each instructor lands on a weighted average of her preferred standard and the
norm. Averaging \eqref{eq:weighted} across the department returns
$\bar g = \E[\hat g_i]$ immediately, because the average of a weighted average of
each instructor and the common mean is that mean again.

\begin{proposition}[Student welfare improves]\label{prop:welfare}
Suppose $F$ is symmetric about its mean or the penalty is quadratic.  Then the intervention strictly reduces the variance of grading standards and leaves
the average unchanged. Student loss therefore strictly falls. Under a quadratic
penalty,
\[
\mathcal{L}(\rho) \;=\; \frac{\kappa}{2}\left(\E[g_i]-\gs\right)^{2}
\;+\; \frac{\kappa}{2}\left(\frac{\lambda}{\lambda+\rho}\right)^{2}\Var(F),
\]
which is strictly decreasing in $\rho$ and tends to
$\tfrac{\kappa}{2}(\E[g_i]-\gs)^{2}$ as the punishment becomes severe.
\end{proposition}

\begin{proof}
By Proposition~\ref{prop:game} every $|d_i|$ shrinks, and by
Proposition~\ref{prop:neutral} the mean is unchanged, so $\Var(g_i)=\E[d_i^2]$
falls. Apply the decomposition of Proposition~\ref{prop:four}: the first term is
unchanged and the second is strictly smaller. The displayed expression follows
from \eqref{eq:weighted}.
\end{proof}

\noindent Proposition~\ref{prop:welfare} is stated for the quadratic $W$ of
Section~3, which was only a second-order approximation. It does not need to be.

\begin{proposition}[The approximation is not doing any work]\label{prop:ssd}
Suppose $F$ is symmetric about its mean or the penalty is quadratic, and let $W$ be \emph{any} concave function. The intervention weakly raises expected
student welfare $\E[W(g_i)]$, strictly if $W$ is strictly concave and the penalty
binds for a positive mass of instructors.
\end{proposition}

\begin{proof}
By Proposition~\ref{prop:game}, $g_i=\bar g+h(\hat g_i-\bar g)$ for a strictly
increasing $h$ with $h(0)=0$ and $|h(x)|<|x|$ for $x\ne0$. So every instructor
moves strictly toward $\bar g$ and none crosses it: the distribution of $g_i$ and
that of $\hat g_i$ therefore cross exactly once, at $\bar g$. By
Proposition~\ref{prop:neutral} they have the same mean. Equal means and a single
crossing are precisely the conditions under which the tighter distribution
second-order stochastically dominates the other, so $\E[W(g_i)]\ge\E[W(\hat g_i)]$
for every concave $W$.
\end{proof}

\noindent In other words the policy produces a mean-preserving contraction of the
distribution of grading standards, and a mean-preserving contraction is good for
students under any concave $W$ --- symmetric or not, quadratic or not, and
regardless of where its peak sits. One caveat deserves stating, since it is easy
to assume the result is more robust than it is: what is needed is
\emph{concavity}, not merely a single peak at $\gs$. A student payoff that is
peaked at the optimum but flattens out in the tails --- so that a section which is
already hopeless is not made much worse by being more hopeless --- is convex far
from $\gs$, and for such a $W$ compression can in principle fail to help. Concavity
says that the marginal damage of a further departure from the optimum is
non-decreasing, which is the substantive assumption.

\noindent The three results together say something worth stating plainly. The
chair can see what her colleagues are doing but not what they ought to be doing,
and a rule written in terms of the departmental average inherits that ignorance:
it cannot move the department toward $\gs$, because nothing in it knows where
$\gs$ is. What it can do is squeeze the instructors together. By
Proposition~\ref{prop:four} this results in an improvement to student welfare.  

\begin{figure}[htbp]
\centering
\includegraphics[width=\textwidth]{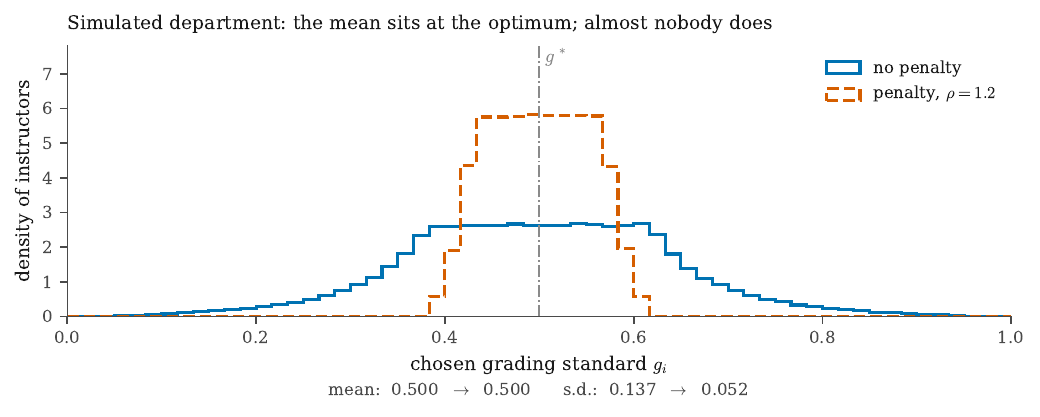}
\caption{A simulated department satisfying Assumption (Interior Grading), in
which bias is symmetric about zero so that average grades sit exactly at the
optimum.  Restraint varies across instructors, independently of bias (see
Appendix~\ref{app:hetero}).  Without the intervention, the mean is $0.500$ and the
standard deviation is $0.137$, with individual sections ranging from three percent
A's to ninety-seven percent.  A quadratic penalty with $\rho=1.2$ leaves the mean at
$0.500$ and reduces the standard deviation to $0.052$.}
\label{fig:department}
\end{figure}

Note also that the intervention cannot push anyone outside $[0,1]$. Every
instructor's standard is a compromise between her own preferred standard and the
norm, and both lie in the interval, so Assumption (Interior Grading) continues to
hold under the policy.  Figure~\ref{fig:department} illustrates the three results in a simulated department.

\subsection{Can the instructors coordinate?}\label{sec:coord}

A natural worry about any rule that punishes deviation from the average is that it creates a coordination game.  If every instructor gives every student an A, no one deviates and no one is punished, and an instructor who holds out for a lower standard is the only one who pays.  This does not happen in our model when the penalty is smooth.  Suppose every other instructor gives every student an A, so that $\bar g=1$.  Under the quadratic penalty, \eqref{eq:weighted} says that instructor $i$ chooses
\[
g_i \;=\; \frac{\lambda\hat g_i+\rho}{\lambda+\rho} \;<\; 1
\]
whenever her own preferred standard is below one.  Her restraint pulls her back toward the standard she would choose on her own, and since every instructor feels this pull, the norm cannot stay at one.  More generally, Proposition~\ref{prop:game} says that the equilibrium is unique, and Proposition~\ref{prop:neutral} says that the norm sits where the instructors' own preferences put it.

Two things are needed for this argument.  The first is net restraint, $\lambda>0$.  It is restraint, not the existence of $\gs$, that anchors the norm.  An instructor with $\lambda=0$ has no preferred standard of her own, and if $\{g_i\}$ is an equilibrium, so is any common shift of it that stays inside $[0,1]$.  Note also that the instructors need not know or agree on $\gs$.  If instructor $i$ believes the optimum is $\gs_i$, this changes only the distribution $F$ of preferred standards, and Propositions~\ref{prop:game} and~\ref{prop:neutral} depend on nothing else.  Disagreement about $\gs$ is simply another source of grading variance.

The second is that the penalty must be mild for small deviations.  If the first unit of deviation is expensive, instructors near the norm stop exactly at it, and when this is true of every instructor, nothing pins the norm down.

\begin{proposition}[When instructors coordinate]\label{prop:coord}
Suppose $\rho_0 \ge \tfrac{\lambda}{2}\left(\hat g_{\max}-\hat g_{\min}\right)$.  Then the equilibria are exactly the profiles in which every instructor chooses the same standard $c$, for any
\[
c\in\left[\hat g_{\max}-\rho_0/\lambda,\;\hat g_{\min}+\rho_0/\lambda\right]\cap[0,1].
\]
In particular, giving every student an A is an equilibrium if and only if $\rho_0\ge\lambda\left(1-\hat g_{\min}\right)$.
\end{proposition}

\begin{proof}
If every other instructor chooses $c$, then $\bar g=c$, and instructor $i$'s best response is $d_i=0$ if and only if $\lambda|\hat g_i-c|\le\rho_0$.  For $c$ in the interval, this holds for every instructor, so the profile is an equilibrium.  If $\bar g>\hat g_{\min}+\rho_0/\lambda$, a positive mass of instructors near $\hat g_{\min}$ choose $d_i<0$, while an instructor with $d_i>0$ would need $\hat g_i>\bar g+\rho_0/\lambda>\hat g_{\min}+2\rho_0/\lambda\ge\hat g_{\max}$, so no instructor does.  Then $\E[d_i]<0$ and $\bar g$ is not an equilibrium.  The case $\bar g<\hat g_{\max}-\rho_0/\lambda$ is symmetric.  The last claim takes $c=1$ and uses $\hat g_{\max}\le1$.
\end{proof}

\noindent A linear penalty, $P(d)=\rho|d|$, has $\rho_0=\rho$, so a sufficiently harsh linear penalty produces exactly the coordination problem described above: any common standard in an interval is an equilibrium, including, if the penalty is harsh enough, every student getting an A.  A rule that is harsh at the margin even for small deviations, such as a sanction for any departure from the average, is therefore the wrong rule.  The fix is to make small deviations cheap, for example by tolerating deviations within a band around the norm.  A penalty that is zero within a band and convex outside it is differentiable at zero, so Propositions~\ref{prop:game} and~\ref{prop:neutral} apply.

\subsection{The shape of the punishment}

Nothing above requires the punishment to be anything more than convex, but the
shape matters for how the burden falls. Under a quadratic penalty every
instructor's deviation shrinks by the same proportion, $\lambda/(\lambda+\rho)$.
A penalty that rises faster than quadratic compresses the instructors furthest
from the norm proportionally more than those near it, which is what a chair
presumably wants.

A flat per-unit fine is the opposite. Because its marginal rate is the same at
every distance, the rate needed to move the instructor furthest out is levied
identically on the instructor barely out of line, and it creates a kink at the
norm that leaves a mass of instructors grading exactly at the average.  If the fine is harsh enough, every instructor bunches and the norm is no longer pinned down (Proposition~\ref{prop:coord}). The
Appendix works out both cases.

\subsection{Disclosure to students}

The Gradient system at NC State allows a student to look up the grade distribution of any course.  Here, we consider the impact of informing students of instructors' distributions.  In this model, there is no penalty for a grading distribution that differs from the departmental average, but instead, we assume it results in a higher $\delta$, the responsiveness
of enrollment to leniency.  

\begin{proposition}[Disclosure to students]\label{prop:seven}
Since $b_i=\alpha_i-\delta\gamma_i$,
\[
\Var(b_i)=\Var(\alpha_i)+\delta^{2}\Var(\gamma_i)-2\delta\,\mathrm{Cov}(\alpha_i,\gamma_i),
\]
which is increasing in $\delta$ whenever $\alpha_i$ and $\gamma_i$ are
uncorrelated. Student-facing disclosure therefore \emph{raises} the dispersion of
grading standards, and with it the variance component of the welfare loss.
\end{proposition}

To emphasize, if students respond positively to an easier grading standard, then informing them of each class's grading distribution increases grading variance and the welfare loss that comes from it.  This is the contrast Prat (2005) draws between disclosing information to a principal and disclosing it to an audience that cannot evaluate the action.

\subsection{Other interventions}

Universities have tried other responses to grading, and their experience fits the
distinctions above.  When Cornell posted median grades online, students moved into
leniently graded courses, and about 45 percent of the subsequent rise in grades
came from this shift (Bar, Kadiyali, and Zussman, 2009).  This is the enrollment
half of Proposition~\ref{prop:seven}: disclosure raised $\delta$, the responsiveness of
enrollment to leniency.  Bar, Kadiyali, and Zussman find no response by the
faculty, however, so the instructor half was not observed at Cornell.  When
Occidental instead gave instructors reports comparing their grades with their
colleagues', previously low-grading courses raised their grades relative to
high-grading ones, which reduced dispersion; the high-grading courses did not
change, so the average did not fall (Lehr, 2016).  Information alone moved only
the low graders, which is one reason our chair imposes a penalty rather than
simply reporting the data.

Other policies target the level.  Wellesley capped the average grade in
introductory and intermediate courses at a B+; high-grading departments complied,
and their enrollments and course evaluations fell (Butcher, McEwan, and
Weerapana, 2014).  Like the binding grade targets in Love and Kotchen (2010), this
is a constraint on the level that binds from above.  A cap or a target requires
the administrator to choose the right level.  The penalty of this section does
not.

\section{Testable implications}

Our model provides several clear, testable predictions.  Here we summarize.

\begin{enumerate}\itemsep2pt
\item \textbf{The lenient tail is junior and non-tenure-track.} Instructors with a high $\alpha_i$ should be over-represented above $\gs$.  These include those on the job market, the untenured, and non-tenure-track faculty whose contracts must be renewed.  Consistent with this, Ahn, Arcidiacono, Hopson, and Thomas (2024) find that tenured and tenure-track faculty prefer lower grades than lecturers.
\item \textbf{The harsh tail is senior and research-active.} Low $\alpha_i$
combined with $\gamma_i>0$ puts tenured, research-strong, large-lecture-averse
faculty below $\gs$.\footnote{Anecdotally, the explanation commonly given for strict grading is ``high standards.''  Our model provides a different explanation for harsh grading.}
\item \textbf{The extremes are the least restrained, not the most biased.} An
instructor's distance from the optimum is $|b_i|/\lambda_i$, which is large when
the pull is large \emph{or} when restraint is small. As a reminder, $\lambda_i=\mu_i-2\theta_i$ is the instructor's conscientiousness (how much she cares about doing the job well) minus twice her cost of grading.  So instructors with extreme grading policies could have strong preferences regarding evaluations and class size, be less conscientious, or find grading more costly than typical.  One implication is that grading support should compress grading in both directions: an instructor with teaching assistants or automatically graded assessments has a lower $\theta_i$, and therefore a higher $\lambda_i$, so she deviates less from $\gs$ whether she leans lenient or harsh.
\item \textbf{A norm enforced by the chair compresses; disclosure to students spreads.}
Propositions~\ref{prop:welfare} and \ref{prop:seven} have opposite signs on
$\Var(g_i)$, which is directly testable wherever one policy was adopted and not
the other.  The faculty side is consistent with Occidental, where reports comparing instructors' grades with their colleagues' reduced dispersion (Lehr, 2016).  The student side has not been shown: at Cornell, students moved toward leniently graded courses, but Bar, Kadiyali, and Zussman (2009) find no response by the faculty.
\item \textbf{A relative-norm policy should compress grading without moving the section average.}  This is the signature of Proposition~\ref{prop:neutral}, and it distinguishes our intervention from an ordinary anti-inflation policy, which moves the mean.

\end{enumerate}

\section{Remarks}

\begin{remark}[What is doing the work]
Two ingredients drive our results: an interior social optimum, and heterogeneity in instructors' bias (which can take
either sign) and restraint.  A third, that grading is cheapest at the extremes, amplifies them: without the cost of effort, the model would still produce two-sided distortions, but they would be uniformly smaller.  Without heterogeneity, our model would predict a uniform shift in grades (inflation if bias is positive) but no dispersion, which is, in effect, what is described in the current literature.
\end{remark}

\begin{remark}[On counting only the students]
The welfare criterion is student welfare throughout, and every welfare statement
in the paper should be read that way. The choice is not innocuous.  In particular, it is what lets Proposition~\ref{prop:welfare} ignore the cost of the penalty, since the chair's sanctions fall on instructors and instructors are not in the
objective.  We think it is a natural assumption, but  Appendix~\ref{app:costly} works out what changes for a reader who
disagrees. The short answer is that a weak penalty can fail to repay its cost, but
a credible one is nearly costless in equilibrium, because deviations, and with them
the sanctions actually imposed, shrink toward zero.
\end{remark}

\begin{remark}[On the common-cohort assumption]
Everything interpretive in our paper rests on this assumption.  If instructors face genuinely different
students, different standards may be optimal, and dispersion is not evidence of
anything.  For example, Calculus I students may differ in the fall semester from the spring.  An online class might have a different student composition than an in-person section, and students who choose an 8:30 section may differ from those who choose a 1:30 section.  The assumption is credible for sections of the same course in the same term and for courses with heavy student overlap, and the model should be taken to
data of that kind.
\end{remark}

\begin{remark}[Section composition and student choice]
No two sections have identical students, and a rule that pulls every section toward the same share of A's will penalize an instructor whose students happen to be stronger.  With a smooth penalty, small differences in composition are cheap, and it is hard to believe that composition explains a difference between 17 and 80 percent A's in two sections of the same course in the same term.  Where composition does differ, a chair can compare each section with the grades its students earn in their other courses rather than with the raw average.  Note also that the rule reduces, rather than creates, the incentive for students to shop among sections.  Today, a student choosing between a section with 17 percent A's and one with 80 percent has a strong reason to choose the second.  Compressing grading variance shrinks that reason.
\end{remark}

\begin{remark}[Common exams and rubrics]
A common final exam, written and graded by someone other than the instructor, holds every section to the same standard, and so does a rubric shared across sections.  Where these are feasible, they address grading variance more directly than our penalty, and we see them as complements rather than alternatives.  However, both require the instructors of a course to agree on coverage and on what an A requires, and a common exam requires someone to write and grade it.  Our rule requires none of this: it uses only data the chair already has.  It also reaches the parts of a grade that a common exam leaves to the instructor, such as homework, projects, and participation.  Note also that neither a common exam nor a rubric chooses the standard.  The department must still decide what an A requires.  Our point is narrower: whatever the standard is, it should be the same across sections.
\end{remark}

\begin{remark}[Courses with a single instructor]
Our rule needs a comparison group, so it does not discipline an instructor who is the only one teaching a course.  Where students overlap heavily, the comparison group can be the other courses that draw on the same students, such as the upper-level electives in a major.  Where there is no such group, our model has nothing to say, and other tools, such as published rubrics, are needed.
\end{remark}


\section{Conclusion}

Grades play two pivotal roles at a university.  They incentivize student effort and signal the student's educational attainment.  Many people have raised concerns along these dimensions about the harm of grade inflation:  when an A is easy, students do not work as hard, and it is difficult for employers to determine who the best students are.  We are raising the same concerns about grading variance.  Students also work less, and consequently learn less, when grades are too harsh.  Further, grading variance degrades the signaling value of a grade, because a grade then reflects not just the performance of the student but also the harshness or leniency of the instructor.

Our model is meant to explain why different instructors choose different grading policies even though they face the same cohort of students and all know the optimal policy.  It provides one explanation for why grading variance is a pervasive problem and also suggests a simple policy intervention.  We should punish instructors who deviate from the departmental mean.

NC State already publishes its grade distributions, yet the variance we describe persists.  This is an informative example of a policy intervention implemented incorrectly.  Currently, the administration does not use the grade distribution to evaluate instructors.  Instead, it is information provided to the students.  As a result, it does not penalize deviation from the mean but rather makes enrollment more responsive to grading leniency.  We have shown that a policy that punishes deviation will not change the department average but instead will reduce the variance.  However, we expect that a policy that does not punish deviation but instead informs students will \emph{increase} the dispersion of standards, because it makes enrollment more responsive to leniency.  Occidental's experience, where grade reports went to the faculty, is consistent with the first half of this argument (Lehr, 2016).  At Cornell, where median grades were posted for students, students moved toward leniently graded courses, although Bar, Kadiyali, and Zussman (2009) find no response by the faculty.

\begingroup

\section*{References}
\small
\noindent\hangindent=1.5em
Achen, A. C. and P. N. Courant (2009). What are grades made of? \emph{Journal of
Economic Perspectives} 23(3), 77--92.

\noindent\hangindent=1.5em
Ahn, T., P. Arcidiacono, A. Hopson, and J. Thomas (2024). Equilibrium grading
policies with implications for female interest in STEM courses.
\emph{Econometrica} 92(3), 849--880.
\href{https://doi.org/10.3982/ECTA17876}{doi:10.3982/ECTA17876}

\noindent\hangindent=1.5em
Babcock, P. (2010). Real costs of nominal grade inflation? New evidence from
student course evaluations. \emph{Economic Inquiry} 48(4), 983--996.
\href{https://doi.org/10.1111/j.1465-7295.2009.00245.x}{doi:10.1111/j.1465-7295.2009.00245.x}

\noindent\hangindent=1.5em
Bar, T., V. Kadiyali, and A. Zussman (2009). Grade information and grade
inflation: The Cornell experiment. \emph{Journal of Economic Perspectives}
23(3), 93--108.
\href{https://doi.org/10.1257/jep.23.3.93}{doi:10.1257/jep.23.3.93}

\noindent\hangindent=1.5em
Bernheim, B. D. (1994). A theory of conformity. \emph{Journal of Political
Economy} 102(5), 841--877.
\href{https://doi.org/10.1086/261957}{doi:10.1086/261957}

\noindent\hangindent=1.5em
Boleslavsky, R. and C. Cotton (2015). Grading standards and education quality.
\emph{American Economic Journal: Microeconomics} 7(2), 248--279.
\href{https://doi.org/10.1257/mic.20130080}{doi:10.1257/mic.20130080}

\noindent\hangindent=1.5em
Bonesr{\o}nning, H. (1999). The variation in teachers' grading practices: Causes
and consequences. \emph{Economics of Education Review} 18, 89--105.
\href{https://doi.org/10.1016/S0272-7757(98)00012-0}{doi:10.1016/S0272-7757(98)00012-0}

\noindent\hangindent=1.5em
Braga, M., M. Paccagnella, and M. Pellizzari (2014). Evaluating students'
evaluations of professors. \emph{Economics of Education Review} 41, 71--88.
\href{https://doi.org/10.1016/j.econedurev.2014.04.002}{doi:10.1016/j.econedurev.2014.04.002}

\noindent\hangindent=1.5em
Butcher, K. F., P. J. McEwan, and A. Weerapana (2014). The effects of an
anti-grade-inflation policy at Wellesley College. \emph{Journal of Economic
Perspectives} 28(3), 189--204.
\href{https://doi.org/10.1257/jep.28.3.189}{doi:10.1257/jep.28.3.189}

\noindent\hangindent=1.5em
Carrell, S. E. and J. E. West (2010). Does professor quality matter? Evidence
from random assignment of students to professors. \emph{Journal of Political
Economy} 118(3), 409--432.
\href{https://doi.org/10.1086/653808}{doi:10.1086/653808}

\noindent\hangindent=1.5em
Chan, W., H. Li, and W. Suen (2007). A signaling theory of grade inflation.
\emph{International Economic Review} 48(3), 1065--1090.
\href{https://doi.org/10.1111/j.1468-2354.2007.00454.x}{doi:10.1111/j.1468-2354.2007.00454.x}

\noindent\hangindent=1.5em
Church, S. A. and C. N. Srivastava (2025). Harvard College's grading system is
``failing,'' report on grade inflation says. \emph{The Harvard Crimson}, October 27.
\url{https://www.thecrimson.com/article/2025/10/27/grading-workload-report/}

\noindent\hangindent=1.5em
Claybaugh, A. (2025). Re-centering academics at Harvard College: Update on
grading and workload. Office of Undergraduate Education, Harvard College,
October.

\noindent\hangindent=1.5em
Correa, H. (2001). A game theoretic analysis of faculty competition and academic
standards. \emph{Higher Education Policy} 14, 175--182.

\noindent\hangindent=1.5em
Costrell, R. M. (1994). A simple model of educational standards.
\emph{American Economic Review} 84(4), 956--971.

\noindent\hangindent=1.5em
Denning, J. T., R. L. Nesbit, N. G. Pope, and M. Warnick (2026). Easy A's, less
pay: The long-term effects of grade inflation. NBER Working Paper 34952.
\href{https://doi.org/10.3386/w34952}{doi:10.3386/w34952}

\noindent\hangindent=1.5em
Dickson, V. A. (1984). An economic model of faculty grading practices.
\emph{Journal of Economic Education} 15, 197--203.

\noindent\hangindent=1.5em
Dubey, P. and J. Geanakoplos (2010). Grading exams: 100, 99, 98, \dots\ or A, B,
C? \emph{Games and Economic Behavior} 69(1), 72--94.
\href{https://doi.org/10.1016/j.geb.2010.02.001}{doi:10.1016/j.geb.2010.02.001}

\noindent\hangindent=1.5em
Fischer, P. and S. Huddart (2008). Optimal contracting with endogenous social
norms. \emph{American Economic Review} 98(4), 1459--1475.
\href{https://doi.org/10.1257/aer.98.4.1459}{doi:10.1257/aer.98.4.1459}

\noindent\hangindent=1.5em
Franz, W.-J. I. (2010). Grade inflation under the threat of students' nuisance:
Theory and evidence. \emph{Economics of Education Review} 29(3), 411--422.
\href{https://doi.org/10.1016/j.econedurev.2009.10.013}{doi:10.1016/j.econedurev.2009.10.013}

\noindent\hangindent=1.5em
Gans, J. S. and S. D. Kominers (2026). What does a grade mean? Informativeness
and strategic manipulation of grading systems. NBER Working Paper 35183.
\href{https://doi.org/10.3386/w35183}{doi:10.3386/w35183}

\noindent\hangindent=1.5em
Gershenson, S., S. B. Holt, and A. Tyner (2024). Making the grade: The effect of
teacher grading standards on student outcomes. \emph{Contemporary Economic
Policy} 42(2), 305--318.
\href{https://doi.org/10.1111/coep.12637}{doi:10.1111/coep.12637}

\noindent\hangindent=1.5em
Glaeser, E. L. and J. A. Scheinkman (2003). Nonmarket interactions. In
M. Dewatripont, L. P. Hansen, and S. J. Turnovsky (eds.), \emph{Advances in
Economics and Econometrics: Theory and Applications, Eighth World Congress},
Vol.~I. Cambridge University Press, 339--370.
\href{https://doi.org/10.1017/CBO9780511610240.012}{doi:10.1017/CBO9780511610240.012}

\noindent\hangindent=1.5em
Green, J. R. and N. L. Stokey (1983). A comparison of tournaments and contracts.
\emph{Journal of Political Economy} 91(3), 349--364.
\href{https://doi.org/10.1086/261153}{doi:10.1086/261153}

\noindent\hangindent=1.5em
Gregor, M. (2021). Electives shopping, grading policies and grading competition.
\emph{Economica} 88(350), 364--398.
\href{https://doi.org/10.1111/ecca.12358}{doi:10.1111/ecca.12358}

\noindent\hangindent=1.5em
Holmstr\"om, B. (1982). Moral hazard in teams. \emph{Bell Journal of Economics}
13(2), 324--340.
\href{https://doi.org/10.2307/3003457}{doi:10.2307/3003457}

\noindent\hangindent=1.5em
Kandel, E. and E. P. Lazear (1992). Peer pressure and partnerships.
\emph{Journal of Political Economy} 100(4), 801--817.
\href{https://doi.org/10.1086/261840}{doi:10.1086/261840}

\noindent\hangindent=1.5em
Lazear, E. P. and S. Rosen (1981). Rank-order tournaments as optimum labor
contracts. \emph{Journal of Political Economy} 89(5), 841--864.
\href{https://doi.org/10.1086/261010}{doi:10.1086/261010}

\noindent\hangindent=1.5em
Lee, S. (2026). Harvard faculty votes to limit A's in undergrad courses.
\emph{WBUR}, May 20.
\url{https://www.wbur.org/news/2026/05/20/harvard-faculty-votes-to-limit-as-in-undergrad-courses}

\noindent\hangindent=1.5em
Lehr, B. (2016). Information and inflation: An analysis of grading behavior.
\emph{B.E. Journal of Economic Analysis \& Policy} 16(2), 755--783.
\href{https://doi.org/10.1515/bejeap-2015-0138}{doi:10.1515/bejeap-2015-0138}

\noindent\hangindent=1.5em
Love, D. A. and M. J. Kotchen (2010). Grades, course evaluations, and academic
incentives. \emph{Eastern Economic Journal} 36(2), 151--163.
\href{https://doi.org/10.1057/eej.2009.6}{doi:10.1057/eej.2009.6}

\noindent\hangindent=1.5em
MacLeod, W. B. (2003). Optimal contracting with subjective evaluation.
\emph{American Economic Review} 93(1), 216--240.
\href{https://doi.org/10.1257/000282803321455232}{doi:10.1257/000282803321455232}

\noindent\hangindent=1.5em
McKenzie, R. and E. Allensworth (2025). Variability in the grades teachers give:
Teacher grade value-add, mismatch, and long-term effects. \emph{American
Educational Research Journal} 62(4), 763--797.
\href{https://doi.org/10.3102/00028312251345094}{doi:10.3102/00028312251345094}

\noindent\hangindent=1.5em
Ostrovsky, M. and M. Schwarz (2010). Information disclosure and unraveling in
matching markets. \emph{American Economic Journal: Microeconomics} 2(2), 34--63.
\href{https://doi.org/10.1257/mic.2.2.34}{doi:10.1257/mic.2.2.34}

\noindent\hangindent=1.5em
Popov, S. V. and D. Bernhardt (2013). University competition, grading standards,
and grade inflation. \emph{Economic Inquiry} 51(3), 1764--1778.
\href{https://doi.org/10.1111/j.1465-7295.2012.00491.x}{doi:10.1111/j.1465-7295.2012.00491.x}

\noindent\hangindent=1.5em
Prat, A. (2005). The wrong kind of transparency. \emph{American Economic Review}
95(3), 862--877.
\href{https://doi.org/10.1257/0002828054201297}{doi:10.1257/0002828054201297}

\noindent\hangindent=1.5em
Prendergast, C. and R. H. Topel (1996). Favoritism in organizations.
\emph{Journal of Political Economy} 104(5), 958--978.
\href{https://doi.org/10.1086/262048}{doi:10.1086/262048}

\noindent\hangindent=1.5em
Sabot, R. and J. Wakeman-Linn (1991). Grade inflation and course choice.
\emph{Journal of Economic Perspectives} 5(1), 159--170.

\noindent\hangindent=1.5em
Shleifer, A. (1985). A theory of yardstick competition. \emph{RAND Journal of
Economics} 16(3), 319--327.
\href{https://doi.org/10.2307/2555560}{doi:10.2307/2555560}

\noindent\hangindent=1.5em
Warnick, M., J. Light, and A. Yim (2026). Instructor value-added in
post-secondary education. Working paper, April 20. SSRN 6616698.
\endgroup

\appendix
\section{Appendix}

In this appendix, we relax several assumptions of the main text.  In
Section~\ref{app:boundary}, we drop Assumption (Interior Grading).  In
Section~\ref{app:sources}, we show that differences in restraint are a source of
grading variance in their own right.  In Section~\ref{app:hetero}, we allow
restraint to vary under the intervention.  In Section~\ref{app:finite}, we consider
a department with a finite number of instructors.  In Section~\ref{app:shape}, we
show how the shape of the punishment matters.  Finally, in
Section~\ref{app:costly}, we consider what changes if we count the instructors'
payoffs in the welfare criterion.

Throughout, we write $R_i \equiv b_i/\lambda_i$ for instructor $i$'s deviation from
the optimum, and $F$ for the distribution of preferred standards
$\hat g_i = \gs + R_i$, as in Section~5.

\subsection{Dropping Interior Grading}\label{app:boundary}

Without Assumption (Interior Grading), the instructor still maximizes a strictly
concave objective on $[0,1]$.  Therefore, she chooses her unconstrained preferred
standard truncated to the interval:
\begin{equation}\label{eq:trunc}
g_i \;=\; \gs + T(R_i),
\qquad
T(x) \;\equiv\; \max\left\{\min\left\{x,\tfrac12\right\},-\tfrac12\right\}.
\end{equation}

\begin{corollary}[Who reaches a boundary]\label{prop:three}
$g_i=1$ if and only if $b_i\ge\lambda_i/2$, and $g_i=0$ if and only if
$b_i\le-\lambda_i/2$.  An instructor gives every student an A, or gives no
student an A, exactly when her private pull is at least half of her net
restraint.
\end{corollary}

\noindent Note that an instructor can reach a boundary either because her pull is
large or because her restraint is small.  The instructors at the endpoints are not
necessarily the most biased; they may simply be the least restrained.

\begin{proposition}[The department under truncation]\label{prop:tobit}
Let $F_R$ be the distribution function of $R_i$, which we take to be continuous.
Then for $x\in(0,1)$,
\[
\Pr\left(g_i\le x\right) \;=\; F_R\!\left(x-\gs\right),
\qquad
\Pr\left(g_i=0\right)=F_R\!\left(-\tfrac12\right),
\qquad
\Pr\left(g_i=1\right)=1-F_R\!\left(\tfrac12\right).
\]
In other words, the observed distribution of grading standards is the distribution
of preferred standards, censored at both endpoints.  The censored instructors form
atoms at $0$ and $1$.
\end{proposition}

\begin{proof}
This follows from \eqref{eq:trunc}, since $T$ is the identity on
$[-\tfrac12,\tfrac12]$ and constant outside it.
\end{proof}

\noindent Dropping the assumption has two consequences.  The first is that our
balance result still holds.

\begin{proposition}[The balance result survives]\label{prop:app-balance}
If the distribution of $R_i$ is symmetric about zero, then $\E[g_i]=\gs$ whether
or not Assumption (Interior Grading) holds.
\end{proposition}

\begin{proof}
$T$ is an odd function, because its two bounds are symmetric about zero.  An odd
function of a random variable that is symmetric about zero is itself symmetric
about zero, so $\E[T(R_i)]=0$.  The expectation exists because $|T|\le\tfrac12$.
\end{proof}

\noindent This is slightly stronger than Corollary~\ref{prop:balanced}.  If we
also dropped the uniform bound in Assumption (Net Restraint), so that $\lambda_i$
could be arbitrarily close to zero, then $R_i$ might not have a finite mean.  The
truncated choice always does.

The second consequence is that the observed variance understates the dispersion of
incentives.

\begin{proposition}[The observed variance understates dispersion]\label{prop:app-var}
$\Var(g_i)\le\Var(\hat g_i)$, with equality if and only if the boundary binds for
a set of instructors of measure zero (or all instructors share the same $R_i$).
\end{proposition}

\begin{proof}
$T$ is a contraction: $|T(x)-T(y)|\le|x-y|$ for all $x,y$, strictly whenever
either argument lies outside $[-\tfrac12,\tfrac12]$.  Apply this to two
independent copies of $R_i$ and take expectations, using
$2\Var(X)=\E\left[(X-X')^2\right]$ for $X'$ an independent copy of $X$.
\end{proof}

\noindent The gap can be large.  $\Var(g_i)$ can never exceed $\tfrac14$, while
$\Var(\hat g_i)$ is large when some instructors have very little restraint.  This is
why we assume interior grading in the main text.  Without it, an observed grade
distribution only gives a lower bound on the dispersion of incentives.  Recovering
the underlying dispersion from censored data is a standard estimation problem.

Finally, the boundaries provide a simple test.  If the distribution of $R_i$ is
symmetric, the two atoms in Proposition~\ref{prop:tobit} have equal mass, since
$F_R(-\tfrac12)=1-F_R(\tfrac12)$.  Therefore, comparing the number of sections
in which everyone gets an A with the number in which no one does tests whether the
pressures in a department are balanced, and it does not require knowing $\gs$.
Note that the test only works in one direction: unequal masses reject symmetry,
but equal masses do not establish that $\E[g_i]=\gs$.

\subsection{Where dispersion comes from}\label{app:sources}

An instructor's deviation depends on both her pull and her restraint.  Here we show
that differences in restraint add to grading variance.

\begin{proposition}[Heterogeneous restraint amplifies dispersion]\label{prop:app-sources}
Suppose Assumption (Interior Grading) holds, $b_i$ and $\lambda_i$ are
independent, and $\E[b_i]=0$.  Then
\[
\Var(g_i) \;=\; \sigma_b^{2}\;\E\!\left[\frac{1}{\lambda_i^{2}}\right]
\;\ge\; \frac{\sigma_b^{2}}{\left(\E[\lambda_i]\right)^{2}},
\]
where $\sigma_b^2=\Var(b_i)$, with equality if and only if $\lambda_i$ is the same
for every instructor.
\end{proposition}

\begin{proof}
By independence, $\E[b_i/\lambda_i]=\E[b_i]\,\E[1/\lambda_i]=0$, so $\E[g_i]=\gs$
and $\Var(g_i)=\E[R_i^2]=\E[b_i^2]\,\E[1/\lambda_i^2]$.  The inequality is
Jensen's inequality applied to the strictly convex function $x\mapsto 1/x^{2}$.
\end{proof}

\noindent To interpret the result, compare two departments with the same
distribution of bias and the same \emph{average} restraint.  In the first, every
instructor has the same restraint.  In the second, restraint varies.  The second
department has more grading variance.  The reason is that restraint is in the
denominator.  An instructor with half the average restraint deviates twice as far,
while an instructor with twice the average restraint deviates only half as far, so
the two do not offset.

This has a practical implication for a chair.  Two departments can face exactly the
same pressure from teaching evaluations and enrollment and still look very
different on Gradient, simply because one has more variation in how much its
instructors care about grading carefully.

\subsection{Heterogeneous restraint under the intervention}\label{app:hetero}

In Section~\ref{sec:intervention}, we assume that every instructor has the same
restraint, $\lambda_i\equiv\lambda$.  Here we allow restraint to vary, and we take
the penalty to be quadratic, $P(d)=\tfrac{\rho}{2}d^2$.  The first-order condition
is still linear, and instructor $i$ chooses
\[
g_i \;=\; w_i\,\hat g_i + (1-w_i)\,\bar g,
\qquad
w_i \;=\; \frac{\lambda_i}{\lambda_i+\rho}.
\]
Averaging across the department gives the norm:
\[
\bar g \;=\; \frac{\E\!\left[w_i\,\hat g_i\right]}{\E\!\left[w_i\right]}.
\]
The norm is a weighted average of preferred standards.  When $\rho=0$, every
weight is one and $\bar g$ is the simple average.  As $\rho$ rises, the weights
shift toward the more restrained instructors, because they move less toward the
norm.  When restraint is common, the weights are equal and the norm does not move,
which is Proposition~\ref{prop:neutral}.

\begin{proposition}[Heterogeneous restraint]\label{prop:app-hetero}
Suppose the penalty is quadratic and the average preferred standard does not depend
on restraint, $\E[\hat g_i\mid\lambda_i]=\E[\hat g_i]$.  This holds, for example,
if bias is mean zero at every level of restraint.  Then for every $\rho>0$, the
intervention leaves the department average unchanged, strictly reduces the variance
of grading standards, and strictly reduces student loss.
\end{proposition}

\begin{proof}
Let $m=\E[\hat g_i]$.  Then $\E[w_i\hat g_i]=\E\!\left[w_i\,\E[\hat g_i\mid\lambda_i]\right]=m\,\E[w_i]$, so $\bar g=m$.  Since $g_i-\bar g=w_i(\hat g_i-\bar g)$,
\[
\Var(g_i)=\E\!\left[w_i^2\,\E\!\left[(\hat g_i-m)^2\mid\lambda_i\right]\right]
<\E\!\left[\E\!\left[(\hat g_i-m)^2\mid\lambda_i\right]\right]=\Var(\hat g_i),
\]
because $w_i<1$ whenever $\rho>0$.  The average is unchanged and the variance is
smaller, so by Proposition~\ref{prop:four} student loss falls.
\end{proof}

\noindent The condition matters.  If restraint is correlated with the direction of
bias, the norm moves toward the preferred standards of the more restrained
instructors, and this can move the department away from $\gs$.  For example,
suppose half of the instructors are lenient and restrained ($b_i=3$,
$\lambda_i=10$, so $\hat g_i=0.8$) and half are harsh and unrestrained
($b_i=-0.1$, $\lambda_i=1$, so $\hat g_i=0.4$).  Without the intervention, the
average is $0.6$ and the loss is $\tfrac{\kappa}{2}(0.05)$.  As $\rho$ grows, every
instructor converges to a norm of $8.4/11\approx0.76$, and the loss rises to about
$\tfrac{\kappa}{2}(0.07)$.  A moderate penalty still helps in this example, but a
severe one leaves students worse off than no penalty at all.

Note that the intervention still uses no information about $\gs$.  The norm can
move when restraint varies, but nothing in the rule moves it toward the optimum.

\subsection{A finite number of instructors}\label{app:finite}

With $n$ instructors rather than a continuum, an instructor's own choice moves the
norm: $\partial(g_i-\bar g)/\partial g_i = 1-\nicefrac{1}{n}$.  The marginal
penalty in \eqref{eq:foc-game} is therefore multiplied by $1-\nicefrac{1}{n}$, and
the results of Section~\ref{sec:intervention} go through with $P'$ replaced by
$(1-\nicefrac{1}{n})P'$.  In the quadratic case, this is the same as replacing
$\rho$ with $\rho(1-\nicefrac{1}{n})$.\footnote{For a non-quadratic penalty, the
neutrality result requires that the department's preferred standards be symmetric
about their mean, which is the finite-department version of the symmetry assumption
in Proposition~\ref{prop:neutral}.}  The conclusions are unchanged; only the
effective strength of the policy differs.

Note that the penalty is weaker in small departments, because an instructor who
deviates also pulls the norm toward herself.  In a department of three, a third of
any deviation is forgiven automatically.  As $n$ grows, we recover the continuum
case.

\subsection{The shape of the punishment}\label{app:shape}

All three of the following points follow from \eqref{eq:foc-game}.

\emph{Proportional compression.}  Write $\hat d_i = \hat g_i - \bar g$ for the
instructor's deviation without the intervention.  Dividing \eqref{eq:foc-game} by
$\lambda d_i$ gives
\[
\frac{d_i}{\hat d_i} \;=\; \left[1+\frac{P'(d_i)}{\lambda d_i}\right]^{-1}.
\]
This ratio is decreasing in $|d_i|$ if and only if $P'(d)/d$ is increasing, that
is, if and only if $P$ rises faster than a quadratic.  For $P(d)=\tfrac{\rho}{k}|d|^{k}$,
this holds exactly when $k>2$.  Under a quadratic penalty, the ratio is the
constant $\lambda/(\lambda+\rho)$, so every instructor's deviation shrinks by the
same proportion.

\emph{Targeting.}  The marginal penalty on an instructor who deviates by $d$ is
$P'(d)$.  A linear penalty, $P(d)=\rho|d|$, has the same marginal rate at every
distance, so the rate needed to move the instructor furthest from the norm is also
imposed on an instructor who is barely out of line.  A strictly convex penalty has
an increasing $P'$, so it can be severe for large deviations and mild for small
ones.

\emph{Bunching.}  Under $P(d)=\rho|d|$, the objective has a kink at $\bar g$, and
every instructor with $\lambda|\hat g_i-\bar g|\le\rho$ chooses exactly $g_i=\bar g$.
The equilibrium distribution therefore has an atom at the norm, and the atom grows
with $\rho$.  Once $\rho\ge\tfrac{\lambda}{2}(\hat g_{\max}-\hat g_{\min})$, every instructor is in the atom and the norm is no longer pinned down (Proposition~\ref{prop:coord}).  A smooth penalty produces no atom.  This is testable.  Any rule with a
threshold, such as ``no more than $X$ percent A's,'' should produce a spike in the
distribution of grading standards at the threshold.

\subsection{Counting the instructors}\label{app:costly}

Our welfare criterion is student welfare.  Under that criterion,
Proposition~\ref{prop:welfare} does not require weighing the cost of the penalty: the chair's sanctions fall on
instructors, and instructors are not in the objective.  Here we consider a reader
who would also count the instructors.  The answer is that a weak penalty can fail
to pay for itself, but a strong one is nearly costless.

\begin{remark}[If punishment is itself costly]\label{rem:costly}
Suppose the sanctions are real costs rather than transfers, and add them to the
loss:
$\mathcal{L}(\rho)=\tfrac{\kappa}{2}\left[(\E[g_i]-\gs)^{2}+\Var(g_i)\right]+\E[P(d_i)]$.
Under a quadratic penalty, $\mathcal{L}'(\rho)$ has the sign of
$\lambda-2\kappa-\rho$.  So if $\lambda>2\kappa$, a weak penalty increases the
loss: it imposes real sanctions but barely reduces dispersion.  The policy does not
pay for itself until $\rho>\lambda(\lambda-2\kappa)/\kappa$.  However, the
sanctions actually imposed are
$\E[P(d_i)]=\rho\lambda^{2}\Var(F)/2(\lambda+\rho)^{2}$, which goes to zero as
$\rho$ grows.  A severe penalty is nearly costless in equilibrium because
instructors respond to it by deviating very little.  This is an argument for a
strong rule, not an argument against having one.
\end{remark}

\noindent Two further points suggest that instructor payoffs should not be counted
at all.  First, an important instructor payoff is the effort saved by grading
carelessly.  That is a private benefit obtained by imposing a cost on students, and
counting it would treat the moral hazard as partly a legitimate social goal.
Second, when the cohort is fixed, the enrollment motive is a transfer between
instructors: a student attracted to one section is a student lost to another.  A
welfare criterion that counted instructors would have to net this out, which
removes the $-\delta\gamma_i$ term from the social objective but leaves it in each
instructor's private objective.  This is exactly the structure of an externality,
and it is why we do not think a chair should weigh it.

\end{document}